\documentstyle[aps,prl,epsfig]{revtex}

\begin{document}

\draft 

\twocolumn[\hsize\textwidth\columnwidth\hsize\csname@twocolumnfalse\endcsname
 
\title{Free-volume kinetic models of granular matter}

\author{Mauro Sellitto}

\address{      Laboratoire de Physique 
                de l'\'Ecole Normale Sup\'erieure de Lyon \\
                46 All\'ee d'Italie, 69007 Lyon, France
}

\author{Jeferson J. Arenzon}
\address{      
                Instituto de F{\'\i}sica, Universidade 
                Federal do Rio Grande do Sul\\ 
                CP 15051, 91501-970 Porto Alegre RS, Brazil
} 
\maketitle

\begin{abstract}
We show that the main dynamical features of granular media can 
be understood by means of simple models of fragile-glass forming 
liquid (Kob and Andersen, {\it Phys. Rev. E} {\bf 48}  4364, 1993) 
provided that gravity alone is taken into  account. 
In such lattice-gas models of cohesionless and frictionless particles, 
the compaction and segregation phenomena appear as purely 
non-equilibrium effects unrelated to the Boltzmann-Gibbs measure which 
in this case is trivial. 
They provide a natural framework in which  slow relaxation phenomena in 
granular and glassy systems can be explained in terms of a common 
microscopic mechanism given by a free-volume kinetic constraint.
\end{abstract}


\twocolumn\vskip.5pc]\narrowtext


Vibrated powders are one of the most interesting examples of 
non-equilibrium systems where excluded-volume interactions are known 
to play a crucial role.
Based on the free-volume notion the analogy with glassy dynamics
(and its limit) was early recognized~\cite{Struik}.
Indeed, the static and dynamic properties of powders seem to be 
even more intriguing than those appearing in 
amorphous systems~\cite{Anita,Chicago,Herrmann,deGennes,Leo}.
A natural question that arises is to what extent a powder can 
be considered as a glass~\cite{Sam,Jorge}. 

In this Letter we show that compaction and segregation phenomena can be 
reproduced within a simple microscopic model of fragile-glass forming 
liquid~\cite{KoAn}, provided that gravity alone is taken into  account. 
The key ingredient of the model is a free-volume {\it kinetic} constraint
involving only a selection of the possible configuration changes 
compatible with detailed balance and the Boltzmann-Gibbs distribution.
Such kinetic constraints are so effective in describing slow relaxation
phenomena that there is no need to invoke either a specific energetic 
interaction or a particular geometry of particles.
Consequently, the ground state structure and the thermodynamics of the model 
turn out to be trivial.
This is the most peculiar feature of this lattice-gas model as compared to 
the previously proposed ones (see, for 
example~\cite{MoPo,CoHe,NiCoHe,Tetris,BaLo}).
It follows that compaction and segregation phenomena are, at least
in this case, purely dynamical effects unrelated to the Boltzmann-Gibbs
measure. 
In particular, we find that the subtle interplay of kinetic constraints
and the gravity driving force can account for several  experimental 
findings such as logarithmic compaction, vibration dependence of the 
asymptotic packing density and segregation 
phenomena~\cite{Knight95,Williams}.
Such effect appear below a well defined vibration amplitude 
corresponding to the dynamical jamming transition of the model.
Of course, the dynamics we consider is not intended to represent 
in a realistic way the microscopic granular dynamics but rather to
provide an effective stochastic description, typically at 
mesoscopic level.


The model we consider was introduced by Kob and Andersen 
as a lattice-gas model of fragile-glass and used to test 
the predictions of the mode-coupling theory~\cite{KoAn}. 
In its original version the system consists of $N$ particles 
on a simple cubic lattice where there can be at most one particle 
per site. 
There is no cohesion energy among particles: 
the Hamiltonian is zero, ${\cal H}=0$.
At each time step a particle and one of its neighbouring sites are chosen 
at random;
the particle can move to the new site if it is empty and
if the particle has less than $\nu$ nearest neighbours before and
after it has moved.
This kinetic rule is time-reversible and therefore the detailed balance is 
satisfied.
At high density the dynamics slows down because the reduced free-volume makes
harder for a particle to meet the constraints.
Above a certain threshold density the particles are so interlocked that 
macroscopic structural rearrangements are no longer possible and the mobility 
steeply falls to zero.
Such a mechanism is enough to provide the basic glassy 
phenomenology~\cite{KoAn,KuPeSe,Se}.

In the following we consider the Kob-Andersen model 
in presence of gravity.
The system is described now by the Hamiltonian
\begin{eqnarray}
{\cal H} & = & m g \sum_{i=1}^N h_i n_i
\label{H}
\end{eqnarray}
where $g$ is the gravity constant, $n_i=0,1$ are the  occupation variables 
of the particles and $h_i$ is their height. 
We set throughout the mass of particles to one, $m=1$.
Particles are confined in a box closed at both ends and with periodic boundary 
condition in the horizontal direction.
We then assume that the random diffusive motion of ``grains'' 
produced by the mechanical vibrations on the box
can be modeled by a thermal bath at temperature $T$:
particles satisfying the kinetic constraints can move according to
the Metropolis rule with probability min$[1,x^{-\Delta h}]$, where 
$\Delta h = \pm 1$ is the vertical displacement in the attempted elementary 
move and $x = \exp(-g/T)$ represents the ``vibration''.
We set the constraint threshold at $\nu=5$. 
The system underlying geometry is that of a body centered cubic lattice.
It gives the advantage of using a single parameter to describe the combined 
effect of vibration and gravity.
In this case the Markov process generated by the kinetic rules is 
irreducible on the full configurations space provided 
that the box height is large enough:
indeed it is always possible to find a path connecting any two 
configurations by letting, e.g., particles expand in the whole box.
Therefore the static properties of the model are those of a non-interacting
lattice-gas in a gravity field and they can be easily computed.
We have investigated by extensive Montecarlo (MC) simulation the dynamical
behavior of this model for a system of height $16  L$
and transverse surface $L^2/2$ with $L=20$, for a number of 
particles $N=16000$.
The runs are carried out over $10^6$ MC time sweeps (MCs) and the 
observables averaged typically over 20 realizations 
of dynamical evolution for each vibration amplitude $x$.
The packing density is computed in the lower $25\%$
of the system.


When particles are subject to gravity and no vibrations, $x=0$,
the model has a single ground state with equilibrium packing density 
$\rho = 1$.
It could be attained, e.g., by letting the particles fall from the 
top of the box one at a time~\cite{Nota2}. 
Therefore we prepare the system in a fluffy initial state by placing the
particles at random  in the upper half of the box and letting
them fall randomly, $x=0$, until they are not able to move anymore. 
We can assume that the mechanically stable state produced by this 
extensive statistical handling of particles represents a 
random loose packed state. 
In this state the average packing density  turns out to be
$\rho_{\rm rlp} \simeq 0.707$. 
Once the system is prepared in a fluffy configuration the vibration 
with amplitude $x$ is turned on.
We used a continuous shaking procedure which is computationally 
more convenient and gives results qualitatively similar to the
discontinuous tapping.
In fig.~\ref{compact}a we plot  the packing density, $\rho(t)$, as a 
function of time for several values of $x$ in the full range of weak 
and strong tapping.
In spite of the relatively high value of $\rho_{\rm rlp}$  and 
the fact that kinetic constraints can be thought to mimic an effective 
short-range repulsion, the packing density slowly increases 
as time goes on and its asymptotic value is an increasing function of 
the vibration amplitude, even for quite high value of $x$, as in typical 
compaction experiments~\cite{Knight95}.
The slow increase of $\rho(t)$ is consistent, in the full range of weak 
and strong tapping, with a function of the form~\cite{Knight95}:
\begin{equation}
\rho(t) = \rho_{\infty} - \frac{\Delta\rho_{\infty}}{1 +
                                B \ln\left( 1 + t/\tau \right)} \,,
\label{eq.log}
\end{equation}
(see fig.~\ref{compact}b).
This inverse logarithmic law was first discovered in the experiments
of Chicago group~\cite{Knight95}, and is well supported by numerical 
simulations~\cite{NiCoHe,Tetris,BaLo}, and analytical 
approaches~\cite{BoGe,Naim98,Brey} essentially based 
on free-volume models.
However, we find that in an intermediate range of vibration a power-law fit
also gives good results, 
while for stronger vibration, $x>0.3$, a modified logarithmic
law works much better; a detailed discussion of these results will be 
presented elsewhere.
We find that the fit parameters, $\rho_{\infty}$, $B$ and $\tau$
all depend on $x$.
In particular, the asymptotic packing density, 
$\rho_{\infty}(x)$, is always smaller than one and it depends smoothly 
on $x$ in agreement with the experimental results~\cite{Knight95}.
There is an optimal vibration amplitude, 
$x_{\rm max} \simeq 0.9$, at which $\rho_{\infty}$  reaches a maximum;
above $x_{\rm max}$ the asymptotic density become closer and closer 
to the equilibrium value (see~fig.~\ref{compact}c).
The knowledge of thermodynamic properties allow us to find the 
point $x_{\rm g}$ above which the dynamical results coincide with the 
equilibrium ones. 
We find $x_{\rm g} \simeq 0.96$ (see~fig.~\ref{compact}c).
This critical vibration amplitude defines the jamming transition of 
the model which is the direct analogue of the glass transition.
Below $x_{\rm g}$ the system is no longer able to equilibrate 
on the experimentally accessible time-scales.
This is essentially due to a {\it kinetic bottleneck}: at high packing 
density the number of paths leading to the equilibrium configurations is 
much smaller than the one leading elsewhere. 
Therefore, even if the dynamics is in principle always able to reach 
equilibrium configurations, actually  for $x < x_{\rm g}$ such 
configurations are not effectively accessible.
We have checked that the usual MC procedure, obtained  by removing the 
kinetic constraints, leads to the correct equilibrium configuration for 
any $x$.
In this case the equilibrium state is easily reached and
the corresponding packing density $\rho_{\rm eq} (x)$ is a decreasing 
functions of $x$.
Paradoxically, $\rho_{\infty}(x)$ and $\rho_{\rm eq} (x)$ exhibit
opposite monotony properties as long as $x < x_{\rm max}$.


We now turn to the behavior of the global density profile.
In fig.~\ref{profile} we can observe that near the bottom wall 
the strong packing effects give rise to highly structured density 
profiles, with exponentially damped oscillations over a length 
scale which decreases with the vibration strength. 
This effect is interestingly similar
to the layering phenomena appearing in simple fluids in confined geometry 
which are usually interpreted in terms of the ``missing neighbours'' 
interaction of the particles near the wall~\cite{Evans}.
In our case the particle-wall interaction  
is purely kinetic: the ``particles'' building up the wall can
affect or not the number of nearest neighbours particles in the
kinetic rule.
We checked that in both cases the effect persists and results
in a more or less dense bottom layer according to the ``attractive'' 
or ``repulsive'' nature of the kinetic particle-wall interaction.
An experimental investigation of this point would be welcome.
The exponential damping of these oscillations could explain
why this fluid-like behavior is hardly observed in the experiments.
A similar effect takes place near the top of the granular packing. 
However, due to the roughness of the free surface the effect is
less pronounced.


We finally consider the behavior of a mixture with different 
kinetic constraints. 
Indeed, size and mass segregation phenomena are another puzzling 
behavior exhibited by powders and several distinct mechanisms 
have been proposed~\cite{Williams,Rosato,ChiConv,Muzzio}. 
Our main concern here is to show that, in their simplest form,  
segregation phenomena also appear in the present model.
To keep things as simple as possible we consider only the case
of a binary mixture with particles of identical mass. 
The only difference between the two components is purely kinetic:
one type of particle is constrained while the 
other one is not; the latter particles can however block the former ones.
The Hamiltonian of the system is given by~(\ref{H}),
and the microscopic reversibility is still satisfied.
In fig.~\ref{segrega} we plot the time evolution of the density profile 
of a binary mixture starting from a random homogeneous distribution of 
particles. 
Although the packing fraction of both components increase, the system 
undergoes, at the same time, a spontaneous unmixing with the constrained 
particles segregating at the top of the mixture (see fig.~\ref{segrega}).  
This turns out to be the case even when the mass of constrained particles 
is bigger than the unconstrained ones, at least in a certain range of the 
mass ratio of the two components.
Such effect can be simply explained in terms of the different mobilities: 
during the shaking process the  unconstrained ``small'' particles 
easily fill the gap beneath the constrained  ``large'' particles whenever 
they  open up; 
this decreases the probability of the slower constrained particles to
fall down.
A similar percolation or sifting mechanism was found in 
ref.~\cite{CaCoHeLoNi}.
In our case it is clearly not related to an equilibrium phase 
transition, since the Hamiltonian is ``blind'' to 
the constraints.


To summarize, we have investigated some aspects of granular dynamics 
in a three dimensional gravity-driven lattice-gas model. 
The key ingredient of the model is a free-volume constraint
implemented in a purely kinetic way~\cite{KoAn}.
Below a critical value of vibration the system undergoes 
a purely dynamical jamming transition.
For weak vibrations the packing density logarithmically increases  
towards an asymptotic value which depends on the vibration amplitude, 
in agreement with experiments~\cite{Knight95}. 
Near the bottom wall and the free surface of the granular cluster
the strong packing effects induce highly structured density profiles;
this phenomenon could be closely related to the one appearing 
in confined liquids. 
When a mixture of particles with different constraints is 
considered a vibration-induced phase-separation appear in the system.
We emphasize that the results mentioned above are obtained in a model 
with trivial thermodynamics without introducing any form of quenched 
randomness neither in the energetic interaction nor in the shape of 
particles.
Therefore all interesting effects have a purely dynamical nature and 
manifest themselves as a departure from the Boltzmann-Gibbs equilibrium 
measure. 
It would be interesting to derive these non-equilibrium features in 
terms of the Edwards' measure~\cite{BaKuLoSe}.

\bigskip

MS is supported by a Marie Curie fellowship of the European Commission 
(contract ERBFMBICT983561). 
JJA is partially supported by the brazilian agency CNPq.


\begin{figure}[f]
\begin{center}
a)
\epsfig{file=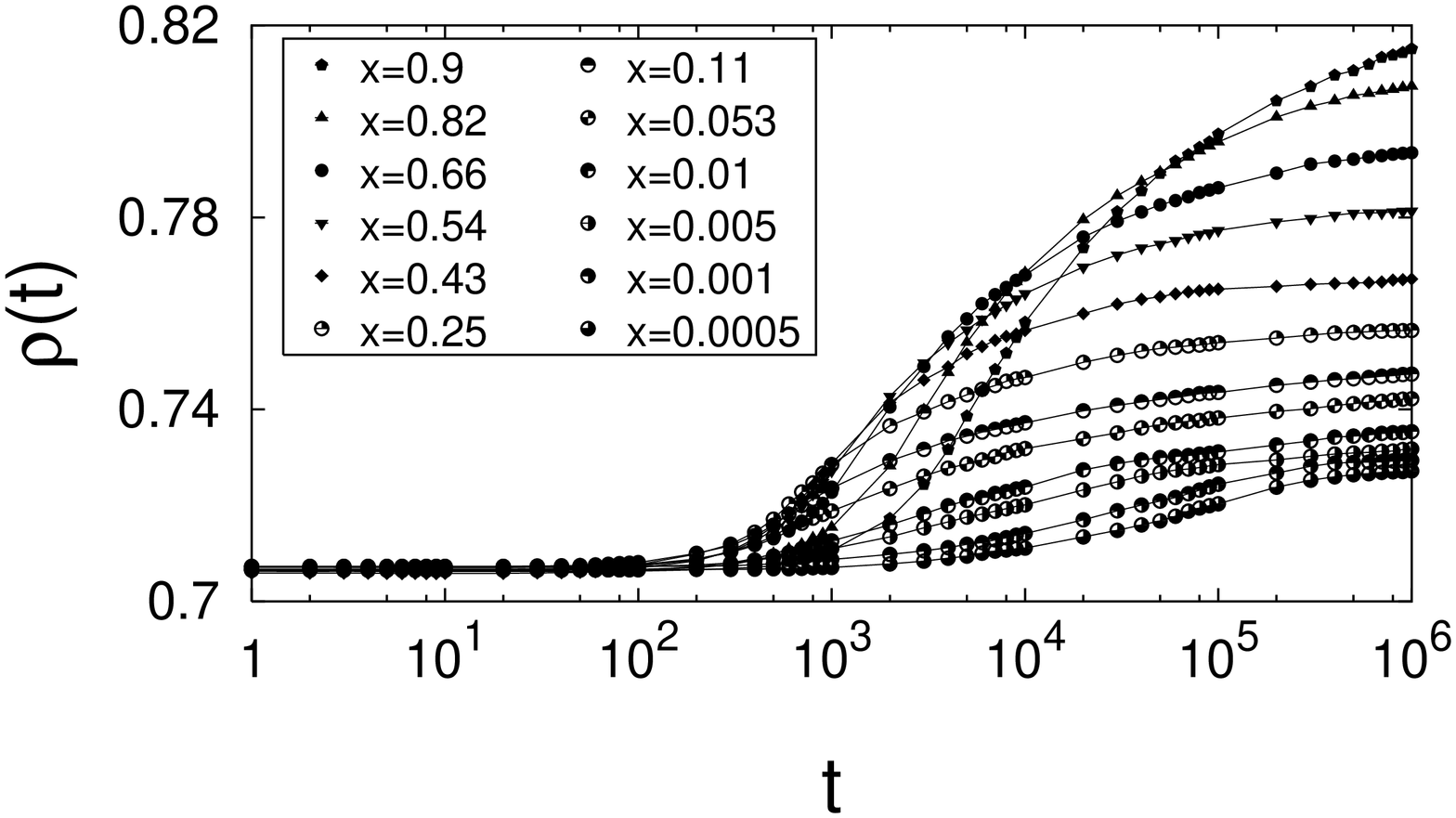,angle=270,width=8cm}
b)
\epsfig{file=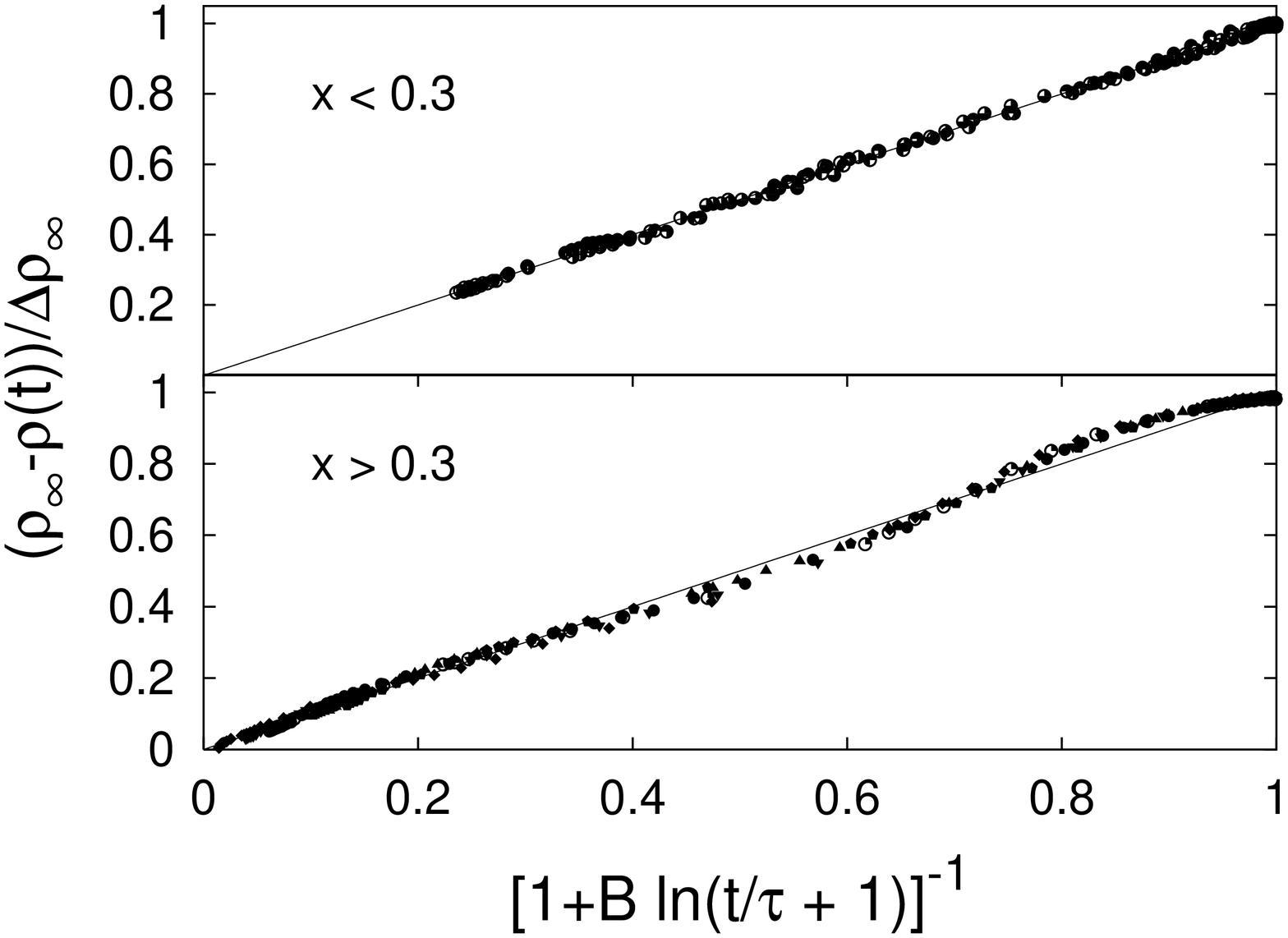,angle=270,width=8cm}
c)
\epsfig{file=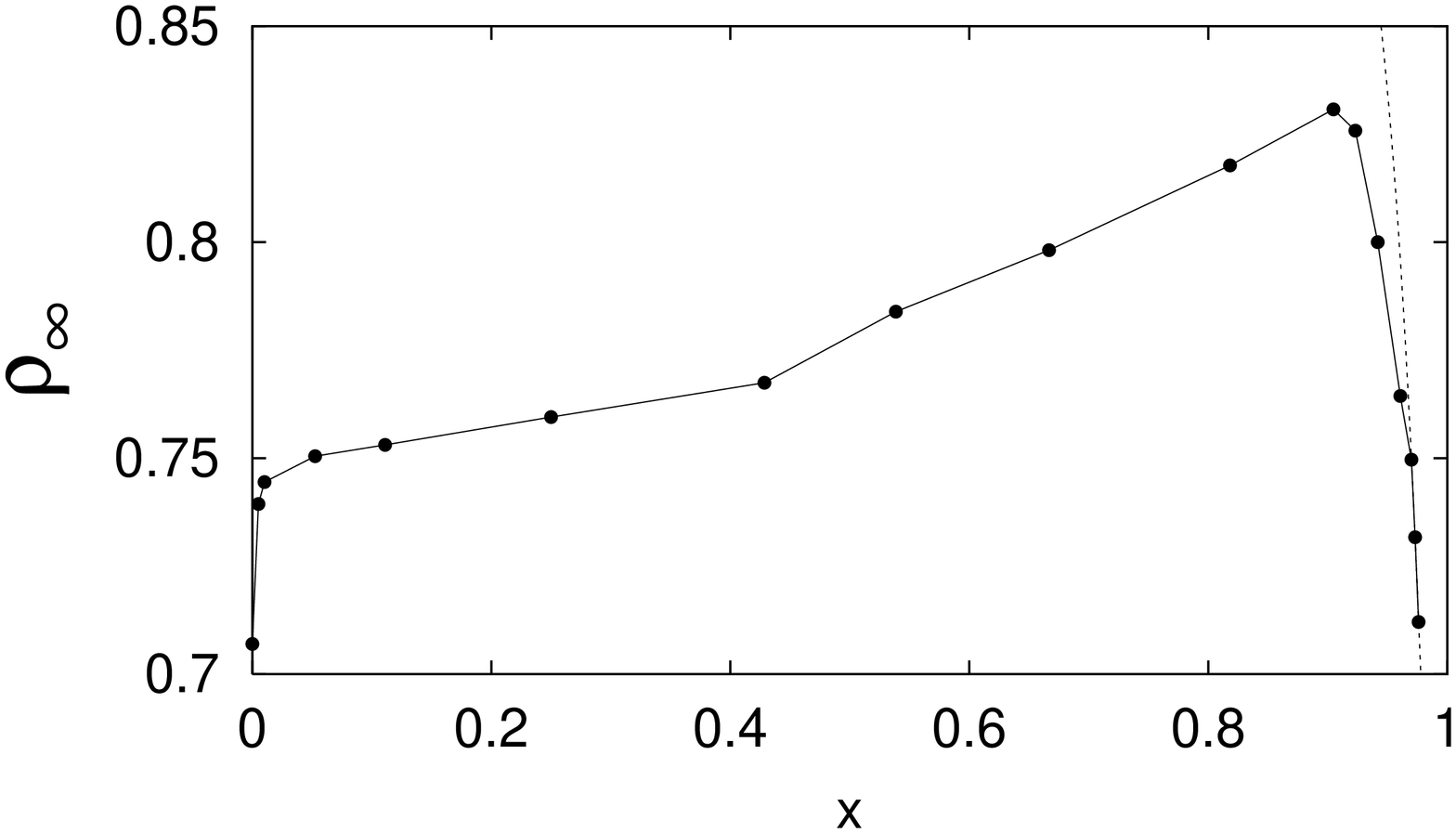,angle=270,width=8cm}
\end{center}
\caption{Compaction experiment data:
a) packing density $\rho(t)$ computed in the lower $25\%$ of the system
   vs time $t$ for several values of the vibration amplitude $x$.
   The initial configuration is prepared in a random loose packed state 
   ($\rho_{\rm rlp} \simeq 0.707$);
b) Logarithmic fit of compaction data in the weak ($x<0.3$) 
   and strong ($x>0.3$) tapping regime. 
   The fit parameters are all dependent on $x$.
c) Asymptotic value of packing density $\rho_{\infty}$ vs~$x$. 
   The maximum occur at $x_{\rm max} \simeq 0.9$ while the
   jamming transition point at  $x_{\rm g} \simeq 0.96$.
   The dashed line represents the exact equilibrium packing density.
}
\label{compact}
\end{figure}

\begin{figure}[f]
\begin{center}
\epsfig{file=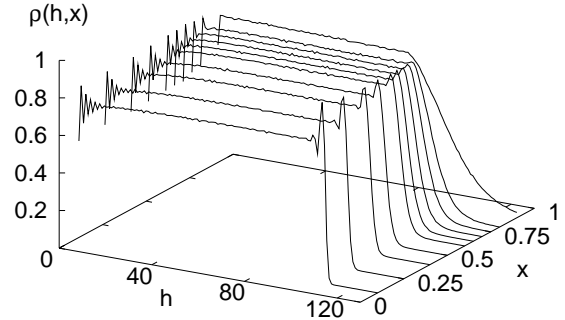,angle=270,width=9cm}
\end{center}
\caption{Density profile $\rho(h,x)$ at time $t=2^{16}$  vs height
$h$ and vibration amplitude $x$.
The bottom wall is located at $h=0$. Here the particle-wall interaction is
``attractive''.}
\label{profile}
\end{figure}

\begin{figure}[f]
\begin{center}
\epsfig{file=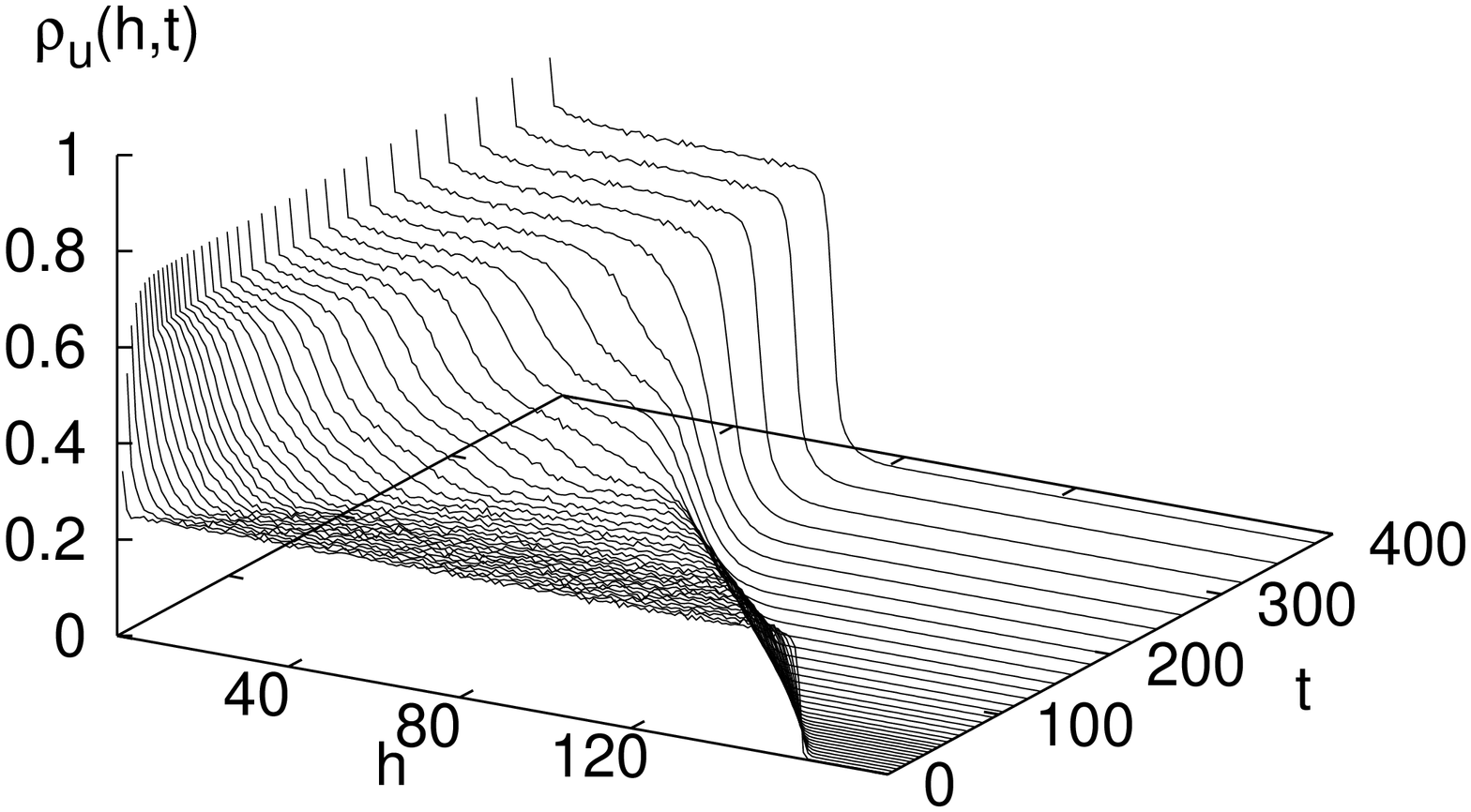,angle=270,width=8cm}
\epsfig{file=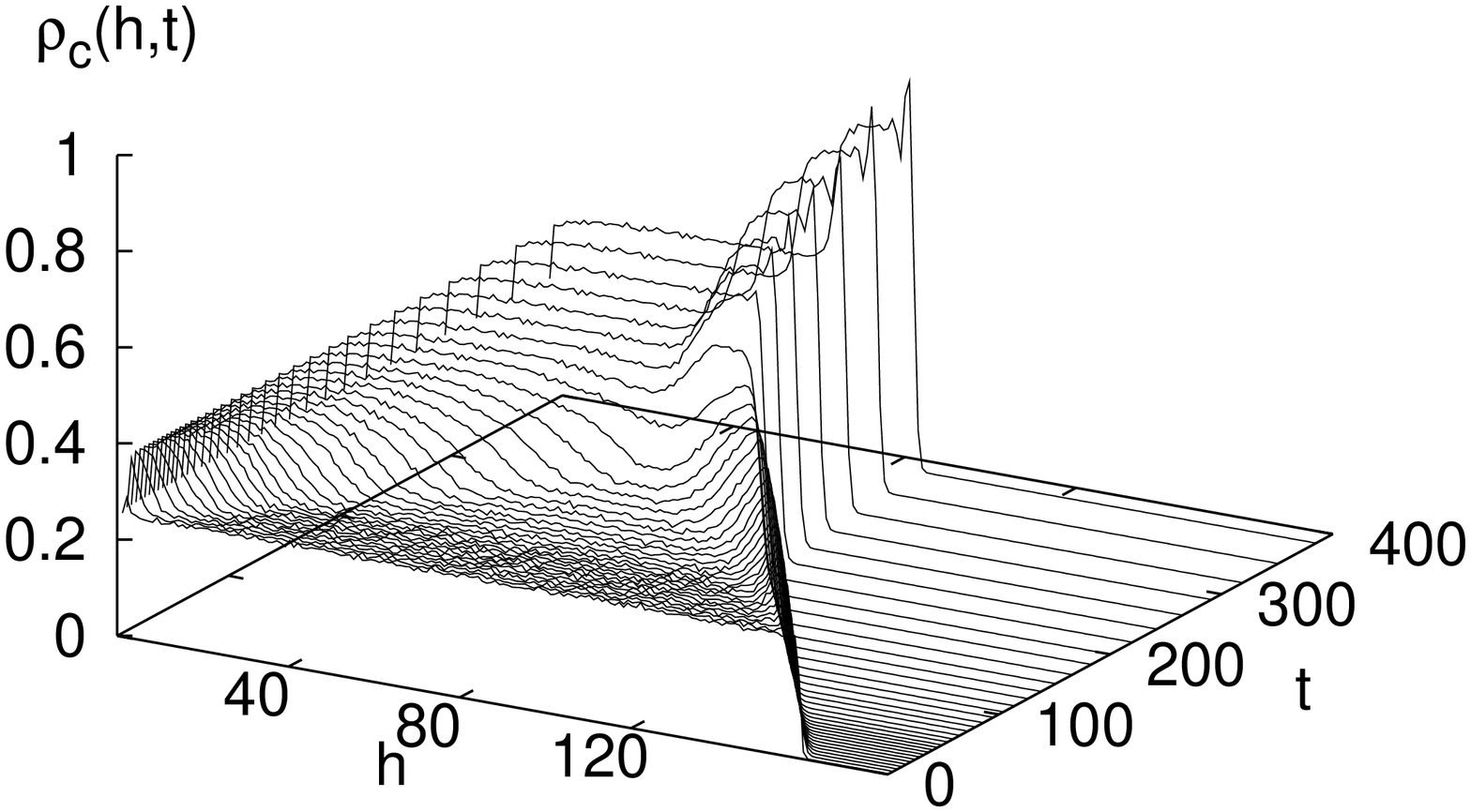,angle=270,width=8cm}
\end{center}
\caption{Density profile of a 1:1 binary mixture vs height $h$ and time $t$.
Top: unconstrained particles; bottom: constrained particles.
The vibration amplitude is $x=0.1$. The system was prepared in a random
homogeneous initial state.
}
\label{segrega}
\end{figure}    
  

\begin{references}
\bibitem{Nota2} This strategy does not work for the simple cubic lattice.
\end{references}
\end{document}